# SEAL: Common Core Libraries and Services for LHC Applications

J. Generowicz, P. Mato, L. Moneta, S. Roiser
*CERN, Geneva, Switzerland*

M. Marino
*LBNL, Berkeley, CA 94720, USA*

L. Tuura
*Northeastern University, Boston, MA 02115 , USA*

The CERN LHC experiments have begun the LHC Computing Grid project in 2001. One of the project's aims is to develop common software infrastructure based on a development vision shared by the participating experiments. The SEAL project will provide common foundation libraries, services and utilities identified by the project's architecture blueprint report. This requires a broad range of functionality that no individual package suitably covers. SEAL thus selects external and experiment-developed packages, integrates them in a coherent whole, develops new code for missing functionality, and provides support to the experiments. We describe the set of basic components identified by the LHC Computing Grid project and thought to be sufficient for development of higher level framework components and specializations. Examples of such components are a plug-in manager, an object dictionary, object whiteboards, an incident or event manager. We present the design and implementation of some of these components and the underlying foundation libraries in some detail.

## 1. THE BLUEPRINT RTAG

The formal process established in the LHC Computing Grid project (LCG) to capture the requirements of the LHC experiments and identify areas of potential common interest among the experiments is by launching a so called Requirements and Technology Assessment Group (RTAG). RTAG 8 was mandated to define the architectural 'blueprint' for LCG applications:
- ? Define the main architectural domains ('collaborating frameworks') of LHC experiments and identify their principal components.
- ? Define the architectural relationships between these 'frameworks' and components, including Grid aspects, identify the main requirements for their inter-communication, and suggest possible first implementations.
- ? Identify the high-level milestones for each domain and provide a first estimate of the effort needed.
- ? Derive a set of requirements for the LCG

The basic idea is that any piece of software developed by any LCG common project must conform to a coherent overall architectural vision. The main goal is to facilitate the integration of LCG and non-LCG software to build coherent applications. The blueprint is established in terms of a set of requirements, suggested approaches and guidelines, and recommendations.

The findings, guidelines and recommendations of the RTAG are summarized in the RTAG report [1]. Here are some of the identified architectural elements:
- ? Interface model with abstract interfaces, versioning and guidelines.
- ? *Component model*. Communication via public interfaces (no hidden channels), plug-ins (run-time loading), life-time management (reference counting), application and component configuration.
- ? *Design guidelines*. Software dependencies, exception handling, interface to external components.
- ? *Object Dictionary*. The ability to query a class about its internal structure (Introspection). Essential for data browsing, rapid prototyping, persistency, etc.
- ? *Object Whiteboard*. Uniform access to application-defined objects (equivalent to the Gaudi transient stores [2]).
- ? *Component Bus.* To easy the integration of components providing a wide variety of functionality and developed independently.

The overall software structure and the role of the different frameworks, ranging from generic ones to specialized to given domain is shown in Figure 1.

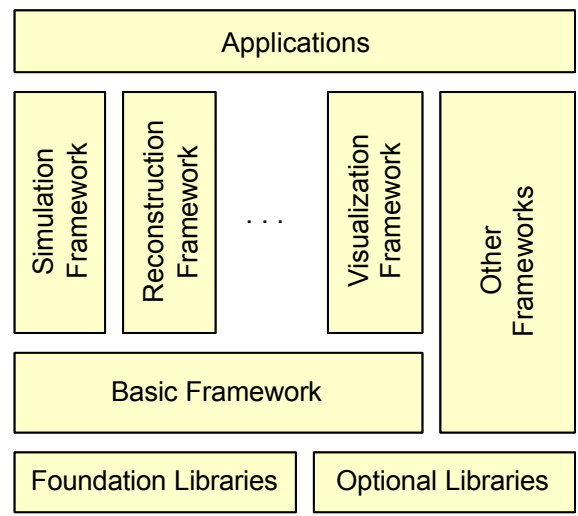

Figure 1 Diagram showing the basic software structure with the different levels from foundations libraries to application software.

The recommendations of the blueprint RTAG can be summarized as:
- ? RTAG establishes a user/provider relationship between LCG software and ROOT [3]. LGC software will not be based on ROOT; it will use ROOT where appropriate.

- Start common project on core tools and services (the SEAL Project)
- Start common project on physics interfaces
- Start an RTAG on analysis, including distributed aspects
- Tool/technology recommendations such as CLHEP, CINT, Python, Qt, and AIDA
- Develop a clear process for adopting third party software.

## 2. THE SEAL PROJECT

The purpose of the SEAL project is to provide the software infrastructure, basic frameworks, libraries and tools that are common among the LHC experiments. The project should address the selection, integration, development and support of foundation and utility class libraries. These utilities cover a broad range of unrelated functionalities and it is essentially impossible to find a unique optimum provider for all of them. They should be developed or adapted as the need arises. In addition to these foundation and utility libraries, the project should develop a coherent set of basic framework services to facilitate the integration of LCG and non-LCG software to build coherent applications.

The scope of the SEAL project covers a big area in the domain decomposition as is shown in Figure 2. The basic two areas are: the *Foundation and Utility libraries*, and the *Basic Framework services*. The Foundation and Utility libraries include the basic types in addition to the ones provided by la programming language (Boost, CLHEP, …), utility libraries, system isolation libraries, domain specific foundation libraries. The Basic Framework Service includes the component model, reflection, plug-in management, incident (event) management, distributed computing, grid, scripting, etc.

In addition to the two main areas, some elements of the interactive and grid services are also included since is very likely that these elements will be common to various LCG projects.

The SEAL project should provide a coherent and as complete as possible set of core classes and services in conformance with overall architectural vision described in the Blueprint RTAG. The following are some assumptions, constraints and risks about the project:

- We will not re-invent wheel. Most of the core software to be delivered by SEAL exists - more or less - in experiments' core software in some form.
- We will re-use as much as possible existing software either from public domain or HEP specific one. Most of the work will be in re-packaging existing pieces of software.
- If what exists is not completely adequate, we will develop / adapt / generalize in order to achieve the necessary level of coherency and conformance to the architectural vision already established in the Blueprint RTAG.
- In order to use SEAL, projects will need to replace their own software elements with SEAL functionally equivalent ones. This will certainly imply some period of instability for the experiment applications.

## 3. WORK PACKGAGES

In the following sections we review the different work packages that have been initiated in the SEAL project. We will be summarizing the current activities and the current

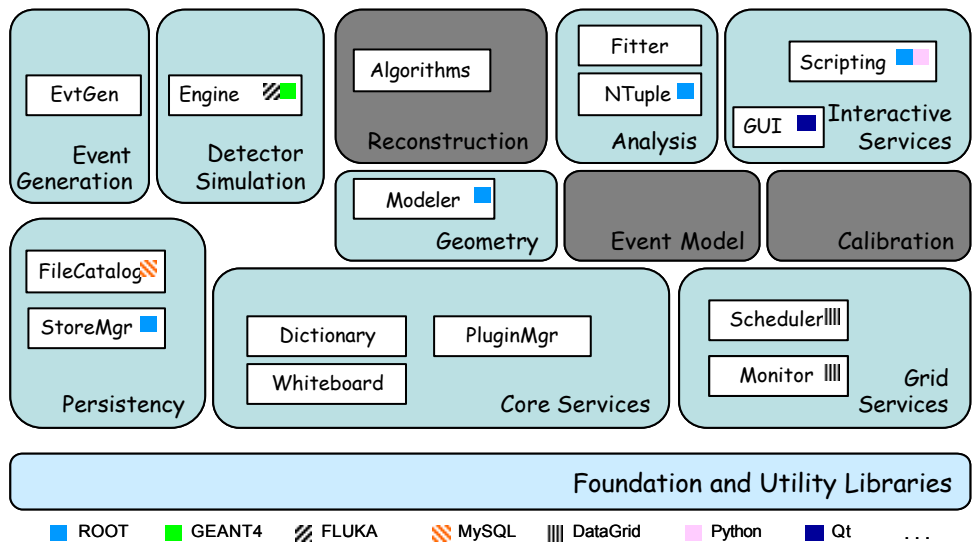

Figure 2 Software Domains covered (light blue) by the applications area of the LCG project and not covered (dark grey).

status and expected deliverables.

## 3.1. Foundation and Utility Libraries

The idea of this work package is to provide class libraries to complement the standard types and utility libraries (a broad range of unrelated functionalities which makes sense to re-use across LCG projects). The goal is to minimize development of foundation and utility libraries in each LCG project and concentrate on the union of the needs in a single set of packages. An inventory of existing utility classes has been produced and the decision to provide support for Boost library [4] has been taken. Boost is an open source utility library, and parts of which are destined to be included in the following C++ standard. Support for CLHEP [5] is also envisaged since it provides HEP specific data types in wide use already in the experiments.

We are developing the SEAL utility and system library with pieces complementary to Boost and STL from existing code from various libraries in use currently in the experiments. The first versions of such libraries have been released and they are in use by other LCG projects.

## 3.2. Math Libraries

The Math Libraries project, originally an independent LCG project launched after the conclusions of an RTAG in reviewing math libraries, has become a work package of the SEAL project. We should provide to the experiments with math and statistics libraries to be used in analysis, reconstruction and simulation.

One of the initial activities has been an evaluation of the GSL [6] library with a view to deploying it fully in the experiments' software. The idea is to standardize on this library where mathematical functionality is concerned and give support to the experiments in using it.

The Minuit [7] minimization package is being rewritten in an object-oriented style, in C++. The idea is to provide all the functionality already existing in the original Fortran version, and to make it easily extensible with more performing algorithms using, for example, dynamically loadable plug-ins. This work includes studies of linear algebra packages, investigating their functionality and performance.

## 3.3. Component Model and Plug-in Manager

In the LCG architecture described in the *Blueprint* RTAG, a plug-in is a logical module that encapsulates the concrete implementation of a given service. The plug-in manager service is responsible for loading, activating, deactivating and unloading plug-ins at run-time. This is a functionally that is desired and in use already by all LHC experiments. The plug-in manager holds cached information about what modules are known, and what plug-ins can be instantiated from them as it is shown in Figure 3. The existing implementation is based on the ideas and code from the Iguana [8] project.

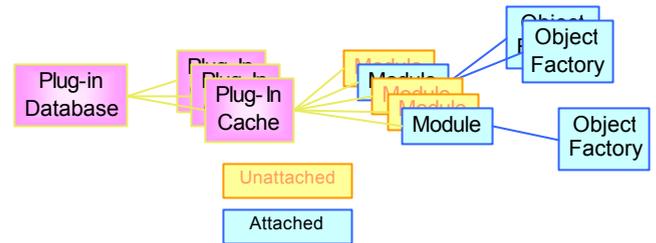

Figure 3 Organization of run time plug-in database provided by the plug-in manager to manage runtime cache information about existing modules and the plug-ins they contain.

In addition to the plug-in manager, we are currently developing the co-called *component model*. This is a set of mechanisms and conventions by which we want to model the components and services that constitute an application. Questions like what shape they have, how the interfaces are exposed and eventually discovered at runtime, identification and configuration, reference counting for implementing an object lifetime strategy, component communication favoring a peer-to-peer strategy are examples of the issues that need to be defined. The aim is to define this component model and provide a number of base classes, interfaces and guidelines that developers can use to develop their own services or applications that comply with the model. The final goal is to achieve an easy way of re-using services and components developed in the different projects and experiments.

## 3.4. LCG Object Dictionary

The *Blueprint* RTAG identified as an essential element the object dictionary to provide reflection functionality by complementing the existing very limited native C++ RTTI (Run Time Type Information). It is highly desirable that this functionality be provided in common across all projects and experiments to exploit the benefits of it. The main idea is to provide pure C++ reflection functionality without any assumption about the possible clients of this information, similar to in the way in which this functionality is provided by other more modern programming languages like Java or C#. Examples of possible clients of the dictionary information are the object persistency layer (e.g. POOL), interactive and scripting services, data browsers, remote communication procedures, etc.

In this work package there are two aspects to be taken into account: the population of the dictionary information from some source and the runtime access to the dictionary information through the reflection interface. The reflection interface is provided by a package that is independent of information source and the client, and provides access to all C++ features.

We have developed tools for populating the dictionary directly from C++ header files, which is the mechanism required by some of the LHC experiments. These tools are based on the gcc_xml [9] package. We are able to generate dictionaries for fairly complex object models without having to instrument or change any input C++ header file. The work flow for the dictionary generation is illustrated in Figure 4. The selection of the classes for which to generate the dictionary and any extra information concerning persistency capabilities or other information is provided by a selection file. The *lcgdict* command uses gcc_xml to generate an intermediate XML representation of a syntax tree describing the contents of C++ header files, which is then used to generate the dictionary filling C++ code (for the "selected classes") from which we produce a dynamic library that can be loaded at run time to create the dictionary in memory.

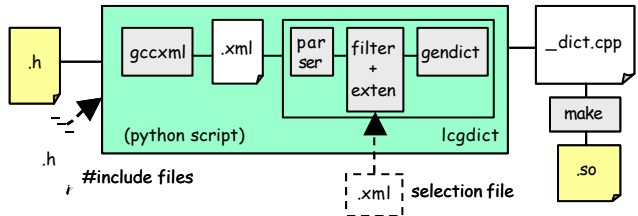

Figure 4 Work flow of the dictionary generation starting from header files using GCC_XML.

As part also of the project, we are currently developing various gateways between different object dictionaries to facilitate the integration and interoperability between languages and frameworks. For example, the way the dictionary is used in the ROOT implementation of the storage manager in POOL [10] is by populating the ROOT dictionary from the LCG dictionary at run-time in the initialization phase, which is then used later directly by the ROOT streaming sub-system. This is shown schematically in Figure 5.

Another gateway that is being developed is a gateway to the Python interpreter. This enables the Python interpreter to access and manipulate any C++ class for which the dictionary information exists, without the need to develop specific bindings for that class. This work also provides a useful completeness check for the LCG reflection interface, as it exercises the behavioral part of the object model which is not used much in the object persistency utilities.

Another required gateway already identified in the *Blueprint* RTAG is the one that would allow a user to interact from ROOT (CINT) with any class in the dictionary as it will be done with Python (inverse direction to the one developed currently in POOL).

### 3.5. Basic Framework Services

In this work package we intend to develop a number of basic framework services based on the component model as described previously. These services should be very basic and neutral to any domain specific framework such as simulation, reconstruction, visualization etc. Examples of these services are: message reporting, exception handling, component configuration, "event" management, object "whiteboard", etc. It is understood that more services of common interest will be identified during the project execution as needs in other projects or in the experiments will arise, and that it will make sense to develop them in common.

The so-called object "whiteboard" is described in the *Blueprint* RTAG report as a mechanism to organize and locate event or detector objects that need to be shared among services or algorithms. This is pattern that appears very often in event data processing applications in our domain, so it makes sense to standardize on something that later can be exploited by adding extra functionality. It is important to study in detail the interaction of this component with the persistency services, visualization services and others.

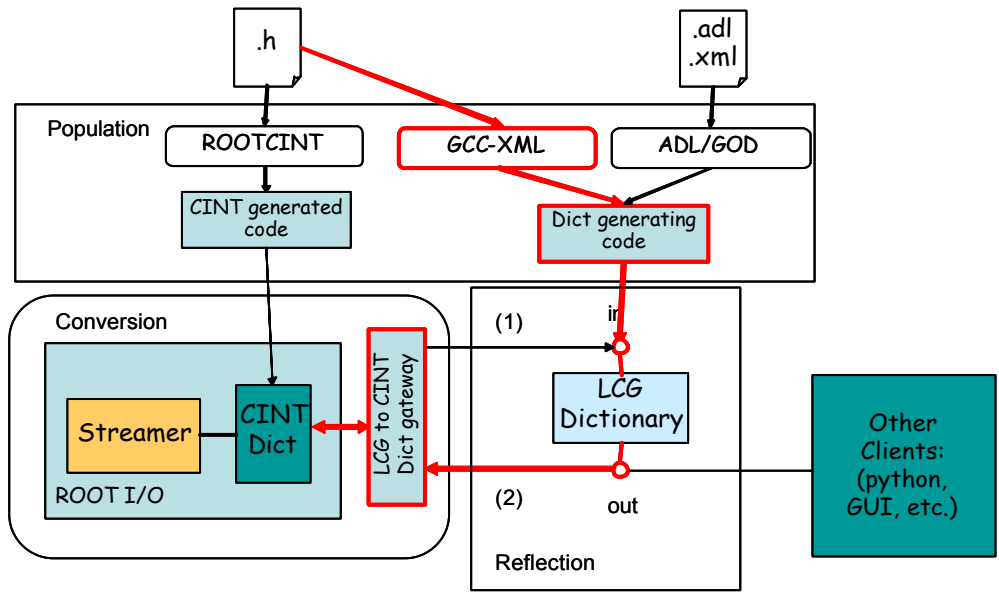

Figure 5 Overall flow of object dictionary information from primary source to its use at run-time by the object persistency layer and other clients.

### 3.6. Scripting Services

The SEAL project should provide the basic infrastructure to support scripting. In particular, two scripting languages (Python and CINT) were identified in the *Blueprint* RTAG. The bindings for Python and CINT of the basic services will be needed to provide a "component bus" that will allow easy integration of components, possibly implemented in a variety of languages and providing a wide range of functionality. Other bindings to more domain specific components will be provided by the corresponding project. For example, bindings for physics analysis tools will be provided in the Physics Interface project.

There is a variety of techniques for developing Python bindings of C++ classes (Python extension modules) in addition the one already mentioned, based directly on using the LCG object dictionary. Therefore, one of the initial activities has been the evaluation of the existing technologies such as SWIG, Boost.Python, SIP, etc. and understanding their interoperability and the possibility of exchanging python objects wrapped using two different technologies. Guidelines for developing Python bindings will be produced in order to guarantee coherency and interoperability between LCG and experiment projects.

Within this work package we will be developing Python bindings to commonly used packages such as CLHEP, GSL, etc. following the agreed guidelines. In the case of ROOT, we have developed already a generic package PyROOT, formerly called RootPython [11], which allows interacting with any ROOT class generically by exploiting the internal ROOT/CINT dictionary.

### 4. STATUS AND CONCLUSIONS

The SEAL project started last November after the recommendations of the *Blueprint* RTAG and has already started to provide the software infrastructure, basic frameworks, libraries and tools that are common among the LCG projects and LHC experiments. Several releases of the project have been made available, which provide some of the functionality that was initially requested. Priority has been given to the pieces required by the POOL persistency project (main emphasis on foundation and utility classes, plug-in management and object dictionary).

A big fraction of the current code base has its origins in various projects across LHC experiments (Iguana, Gaudi, HepUtilities, etc.). This has allowed us to very quickly produce high quality software and with the functionality that we believe is close to what is really needed, since it is based on recent past experiences.

The first more completed release is scheduled for the end of June and should incorporate sufficient functionality to be used by any other LCG project and by LHC experiments' frameworks, to replace their existing equivalent functionality.

### Acknowledgments

The authors wish to thank all people currently participating to the applications area of the LCG project for their contribution in defining the requirements of SEAL project and their support. Specially, we wish to tank the members of the SPI project for their help in setting up the software development infrastructure.